\documentclass[prb,preprint,floatfix]{revtex4-1} 
\usepackage{amsmath} 
\usepackage{amsfonts} 
\usepackage{graphicx} 
\usepackage{braket}

\usepackage[dvipsnames]{xcolor}
\begin{document}

\title{Geometric Visualizations of Single and Entangled Qubits}
\author{Li-Heng Henry Chang, Shea Roccaforte, Ziyu Xu, and Paul Cadden-Zimansky}
\email{paulcz@bard.edu} 
\affiliation{Physics Program, Bard College, Annandale-on-Hudson, NY 12504}

\begin{abstract} 
The Bloch Sphere visualization of the possible states of a single qubit has proved a useful pedagogical and conceptual tool as a one-to-one map between qubit states and points in a 3-D space. However, understanding many important concepts of quantum mechanics, such as entanglement, requires developing intuitions about states with a minimum of two qubits, which map one-to-one to unvisualizable spaces of 6 dimensions and higher. In this paper we circumvent this visualization issue by creating maps of subspaces of 1- and 2-qubit systems that quantitatively and qualitatively encode properties of these states in their geometries. For the 1-qubit case, the subspace approach allows one to visualize how mixed states relate to different choices of measurement in a basis-independent way and how to read off the entries in a density matrix representation of these states from lengths in a simple diagram. For the 2-qubit case, a toroidal map of 2-qubit states illuminates the non-trivial topology of the state space while allowing one to simultaneously read off, in distances and angles, the level of entanglement in the 2-qubit state and the mixed-state properties of its constituent qubits. By encoding states and their evolutions through quantum logic gates with little to no need of mathematical formalism, these maps may prove particularly useful for understanding fundamental concepts of quantum mechanics and quantum information at the introductory level. 
\end{abstract}

\maketitle 

\section{Introduction} 
The conventional visualization model for the states of a single qubit is a Bloch Sphere, which provides a one-to-one correspondence between points in a visualizable geometric space and the quantum mechanical state space. However, many of the interesting and unintuitive properties of quantum mechanics occur only in multi-qubit systems. One of the shortcomings of the conventional Bloch Sphere is that it characterizes only individual qubits rather than correlations between them that dominate the state space of multi-qubit systems. The increasing dimensionality of parameters needed to describe an $n$-qubit system grows exponentially with $n$, making visualizations of even entangled 2-qubit states an intrinsically hard problem. Nevertheless, the conceptual and pedagogical utility of visualizable maps and the growing interest in the study and teaching of quantum information concepts has informed the development of several approaches to visualize $n\geq{2}$ qubit states, Hilbert spaces, and entanglement \cite{charEnt, geoEntBellIneq, onGeoEnt, geoQuantState, geoEnt}.  Strategies to visualize 2-qubit systems include a tetrahedron nested in an octahedron to map equivalence classes of different states \cite{2qubitViz}, two 1-qubit Bloch spheres that track the 2-qubit correlations using color-map ellipsoids \cite{mqubitViz}, embedded steering ellipsoids \cite{SteerEllipse}, or embedded axes \cite{entBloch}, and separate vectors on three-sphere maps that characterize the two 1-qubit states and their entanglement \cite{23qHopf, 2qBS}. Past works have established visual models that are useful for capturing the fullest extent of the 2-qubit state space, albeit using complicated parameterizations that require intimate knowledge of the quantum formalism.

Our approach to visualizing single and entangled qubit states takes its cue from 2- and 3-D Minkowski diagrams, reduced-dimensional cross-sections of the full 4-D space-time map that preserve many of its properties. Mathematically, our cross-sections of Hilbert spaces correspond to states with representations specified by only real numbers, which has the added advantage that one can choose to introduce the corresponding basic linear algebra formalism for the mapped states without the need for complex numbers. Geometrically, the real cross-section of a 1-qubit state is mapped to a circular cross-section of the familiar Bloch sphere, a “Bloch Circle”, while the real cross-section of all pure 2-qubit states is mapped to a solid annular toroid (a donut with a donut-shaped hole \cite{knivesout}). 
The toroidal geometry is chosen to simultaneously allow the encoding of the degree of entanglement of the 2-qubit state in a single length and the states of the underlying 1-qubit mixed states in lengths and angles. Moreover, the geometry demonstrates the difference in abundance of separable and entangled states, illuminates the non-trivial topology of the full 2-qubit state-space, and builds intuition for why maximally entangled 2-qubit states are so frequently employed in basic quantum information algorithms. Thus, our toroid model can serve as an intuitive tool for teaching and learning qubit concepts at an introductory level.  At a more advanced level, the 1-qubit and 2-qubit maps can track the evolution of quantum states in the Hilbert subspaces as qubits are manipulated by quantum logic gates (see Fig.~\ref{figLogicGates}).

To construct our 2-qubit toroid visualization, we first review some under-appreciated geometrical properties of the 1-qubit Bloch Sphere model by analyzing a Bloch Circle cross-section of it. A particular point of emphasis in this review is understanding how to visualize 1-qubit mixed quantum states and the associated entries in the density matrices that represent them.  Due to the increased mathematical complexity of performing algebraic calculations with density matrices, the introduction of mixed states is often postponed to more advanced courses or omitted entirely in undergraduate eduction, despite the fact that almost all physical systems are properly described by mixed states rather than pure ones -- for example the polarization state of photons in the partially polarized or unpolarized light from most light sources.  By emphasizing how one can directly see the measurement probabilities of all possible 1-qubit measurements on any mixed state plotted in a Bloch Circle and read off the density matrix elements for any basis, this model allows the early introduction of mixed states to students without recourse to heavier mathematical formalisms.  Moving to the 2-qubit case, we concatenate two Bloch Circles to construct a torus model that maps separable 2-qubit pure states. The separable torus is extended inward and outward by a finite amount to create a toroidal volume that maps 2-qubit entangled states. The toroid model straightforwardly displays both the 2-qubit state of a system and the corresponding mixed states of each of the constituent qubits, while clearly delineating the 2-qubit states corresponding to separable, entangled, and maximally entangled qubits. This model thus helps learners build intuitions about 2-qubit separable and entangled states, unitary evolutions between them, the corresponding evolutions of the underlying 1-qubit pure and mixed states, and how the relative ``size'' of a state space grows with the number of qubits.
\\
\\
\\
\\
\begin{figure}[ht!]
\centering
\includegraphics[scale=.21]{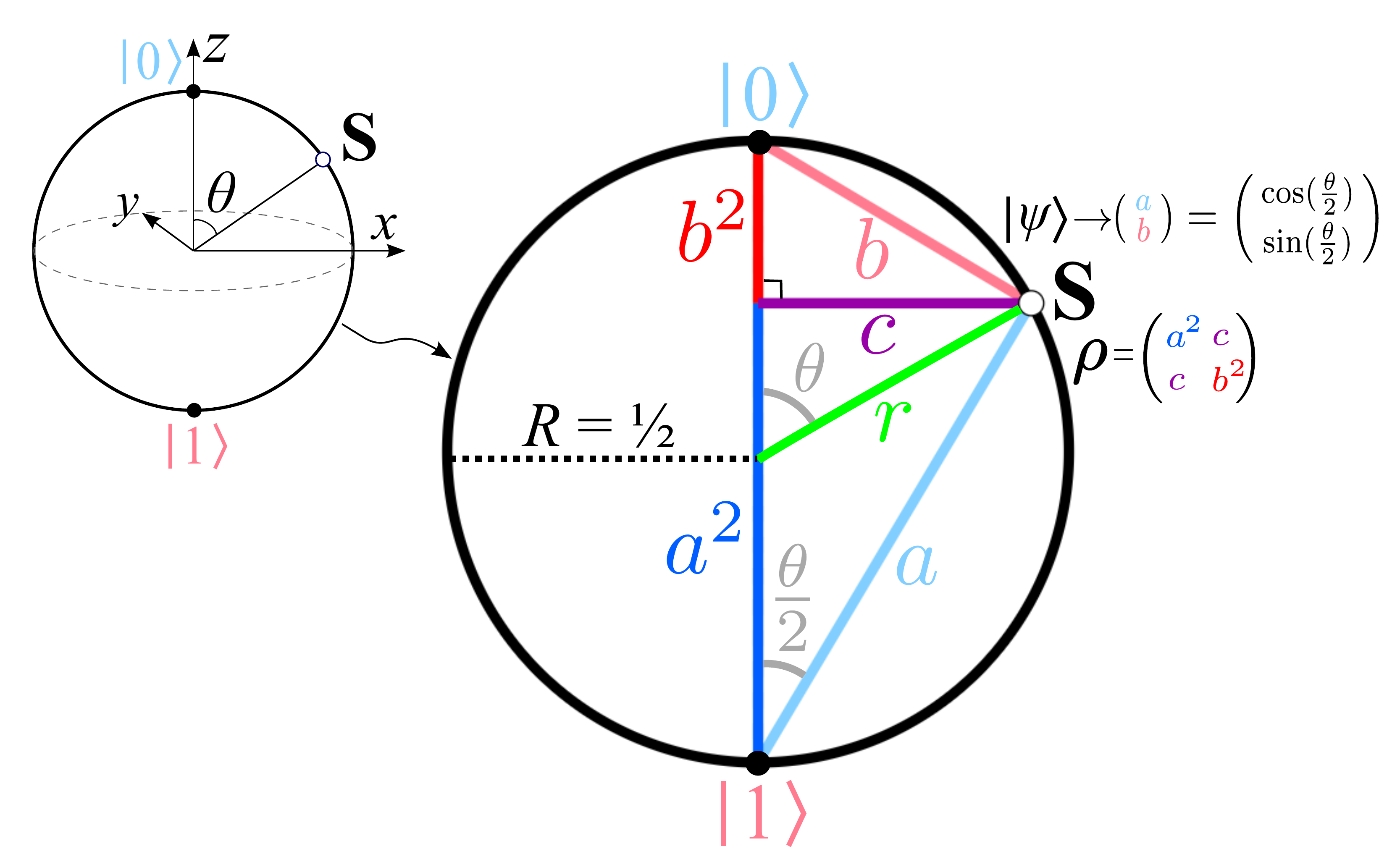}
\caption{A Bloch Circle obtained by taking a cross-section of the conventional Bloch Sphere (upper left corner). The statepoint $S$ at radius $r=R$ on the perimeter of a circle corresponds to a pure state $\ket{\psi}=a\ket{0} + b\ket{1}$. By choosing the radius of the Bloch Circle to be $R={1}/{2}$, the probability amplitudes $a$ and $b$ correspond to the lengths of the lines drawn from the statepoint to the basis statepoints $\ket{0}$ and $\ket{1}$. The perpendicular dropped from the statepoint to the diameter joining the basis points cuts this diameter into two segments with lengths equal to the probabilities $a^2$ and $b^2$ and has a length equal to the magnitude of the off-diagonal element $c$ of the density matrix $\rho$ representing this state.}
\label{figBlochCircle}
\end{figure}
\section{Single Qubits and Bloch Circle Geometry}
The Poincar\'e or Bloch Sphere provides a canonical map with 1-to-1 correspondence between 1-qubit states (up to a global phase uncertainty) and points within a sphere, ``statepoints.'' Pure states composed of a superposition of an orthogonal basis $\ket{0}$ and $\ket{1}$ are represented by column vectors \begin{equation}
\ket{\psi}=a\ket{0} + b\ket{1} \rightarrow \begin{pmatrix} a \\ b \end{pmatrix}=\begin{pmatrix} \cos(\frac{\theta}{2}) \\ \sin(\frac{\theta}{2})e^{i\phi} \end{pmatrix},
\label{eq:pure_statevector}
\end{equation} with the parametrization terms generally introduced as a mix of algebraic, probabilistic, and geometric significance: the $a$ and $b$ probability amplitudes are coefficients of the linear expansion corresponding to the respective probabilities $|a|^2$ and $|b|^2$ of obtaining states $\ket{0}$ and $\ket{1}$ on measurement, while $\theta$ and $\phi$ are the polar and azimuthal coordinates of the statepoint on the surface of the sphere.
Our initial task is to give all these parameters geometrical significance by reviewing the properties of the Bloch Sphere (Fig.~\ref{figBlochCircle}). This task is accomplished first by taking the radius of the Bloch Sphere to be $R=\tfrac{1}{2}$ as opposed to the more common textbook choice\cite{QCtextbook,Schumacher} of an $R=1$ unit sphere.  This choice allows for easily constructed and visualized lines on this modified Bloch Sphere to be exactly equal in length to the magnitudes of state probabilities and probability amplitudes, rather than double their length as they are on a unit sphere.  Second, we focus only on the subspace of states corresponding to $\phi=0$, i.e. a cross-section of the sphere represented by real values of $a$ and $b$ that we term a Bloch Circle. This dimensional reduction makes clear the plane geometric relations between magnitudes and angles and allows for pure 1-qubit quantum states to be specified by a single angle $\theta$ ranging from $-\pi$ to $\pi$. With these choices the parametrization angle $\frac{\theta}{2}$ and lengths corresponding to the magnitudes $a=\cos(\frac{\theta}{2})$, $b=\sin(\frac{\theta}{2})$, $|a|^2$, and $|b|^2$ are easily related to a statepoint located at the perimeter of the circle and the choice of ``measurement'' axis corresponding to the representation (here through the antipodal $\ket{0}$ and $\ket{1}$ states) as shown in Fig.~\ref{figBlochCircle}. We note one virtue of this geometrical diagram is that it is basis independent, so that the corresponding column vector representations and measurement probabilities for a different choice of basis can be intuited geometrically rather than algebraically. 
Proofs of the geometrical properties are found in Appendix \ref{proofGeoProp}.

\begin{figure}[ht!]
\centering
\includegraphics[scale=.35]{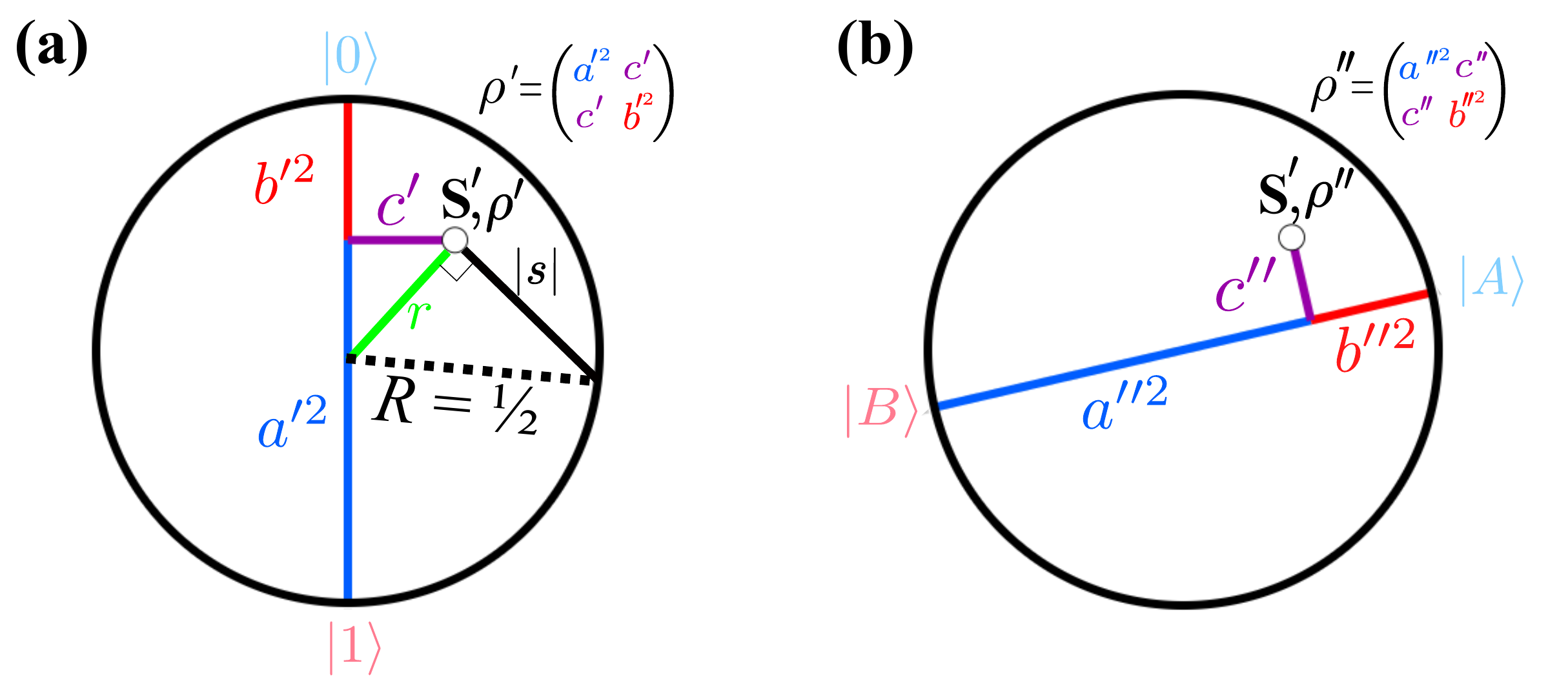}
\caption{(a) The density matrix $\rho'$ for a mixed state plotted by the statepoint $S'$ at radius $r<{R}$ is constructed similarly to the pure-state density matrix of Fig.~\ref{figBlochCircle}. The degree of increasing mixedness of this state compared to a pure state is reflected by the shortening length of the radius $r$ (green) and the corresponding increasing length of the line segment $|s|$ drawn perpendicularly to $r$ from the statepoint to the Bloch Circle circumference, with $|s|=0$ a pure state and $|s|={1}/{2}$ a maximally mixed state. (b) The elements of the density matrix $\rho''$ for the identical state and statepoint $S'$ in a different orthonormal basis $\ket{A}$ and $\ket{B}$ can be similarly constructed using the diameter corresponding to the new basis.}
\label{figMixedStates}
\end{figure}

Another useful representation of the pure 1-qubit state is the density matrix \begin{equation}
\ket{\psi} \bra{\psi} \rightarrow \begin{pmatrix} a \ b \end{pmatrix} \begin{pmatrix} a \\ b \end{pmatrix} = \begin{pmatrix} |a|^2  & ab^* \\  a^*b & |b|^2 \end{pmatrix} = \begin{pmatrix} |a|^2  & c^* \\  c & |b|^2 \end{pmatrix} = \rho,
\label{eq:mixed_matrix}
\end{equation}
where $c\equiv{a^*b}$.  Again, by looking at the subspace of states on the Bloch Circle when all entries are real,
\begin{equation}
\rho = \begin{pmatrix} {\color{blue} a^2}  & {\color{Plum}{c}} \\ {\color{Plum}{c} } & {\color{red} b^2} \end{pmatrix},
\label{eq:rho_color}
\end{equation}
with each entry corresponding to lengths of line segments shown in Figs. \ref{figBlochCircle} and \ref{figMixedStates}, and the off-diagonal matrix element $c$ positive for statepoints on the right-hand side of the Bloch Circle measurement axis and negative on the left-hand side.
Density matrix representations are often omitted from introductions to qubit state descriptions since their algebraic formalism is viewed as more complicated than column vector representations.  However the geometric interpretation and construction of the density matrix entries shown here is quite straightforward and has the added benefit of extending to mixed quantum states, not representable as normalized column vectors, such as those that occur when a qubit is entangled or is viewed as a linear combination of pure qubit states weighted by epistemic, statistical, or classical probabilities.\cite{Schumacher} Indeed, on the road to introducing the concept of entanglement, it is advisable to first acclimate students to mixed states and their density matrix representations, since the constituent particles of an entangled system are invariably described by them.  The general mixed-state matrices have the form
\begin{equation}
\rho = \begin{pmatrix} |a|^2 & c^* \\ c & |b|^2 \end{pmatrix}
\label{eq:rho_complex}
\end{equation}
 with $|c|\leq|ab|$. There is a one-to-one correspondence between these states and statepoints in the interior of the Bloch Sphere, with the $\phi$ argument of $c=|c|e^{i\phi}$ determining the azimuthal angle of the statepoint. Again, as shown in Fig.~\ref{figMixedStates}, we restrict our visualization to the real subspace ($\phi=0$) of points within the Bloch Sphere corresponding to density matrices \begin{equation}
\rho = \begin{pmatrix} {\color{blue} a^2}  & {\color{Plum}{c}} \\ {\color{Plum}{c} } & {\color{red} b^2} \end{pmatrix},
\label{eq:rho_color2}
\end{equation}
which have analogous correspondence between line segment lengths, matrix parameters, and physical significance as the Eq.~\ref{eq:rho_color} and Fig.~\ref{figBlochCircle} pure-state case.  Proofs of these relations are found in Appendix \ref{proofGeoProp}. The degree to which a state is mixed can be visualized geometrically either by the length of the radius $r$ from the center of the Bloch Circle to the statepoint (which monotonically increases with increasing purity from 0 to ${1}/{2}$) or the line segment of length $|s|= \sqrt{\left({1}/{2}\right)^2-r^2}$ drawn from the statepoint to the Bloch Circle perimeter perpendicular to the radius (which monotonically increases with increasing mixedness from 0 to ${1}/{2}$). Below we explain why $s$ is a useful parameter in visualizing 2-qubit states (see also Appendix \ref{proofSameRadius}). As with pure-state representations, the geometrical lengths can be used to read off the density matrix representation of a state for any choice of basis, see Fig.~\ref{figMixedStates}(b).  With this review of the Bloch Sphere and Circle geometry, we can now move to a visualization of 2-qubit states that incorporates features that track the individual qubit states and their degree of entanglement.
\section{Toroid Model for a 2-qubit system}
\begin{figure*}[ht!]
\centering
\includegraphics[scale=.365]{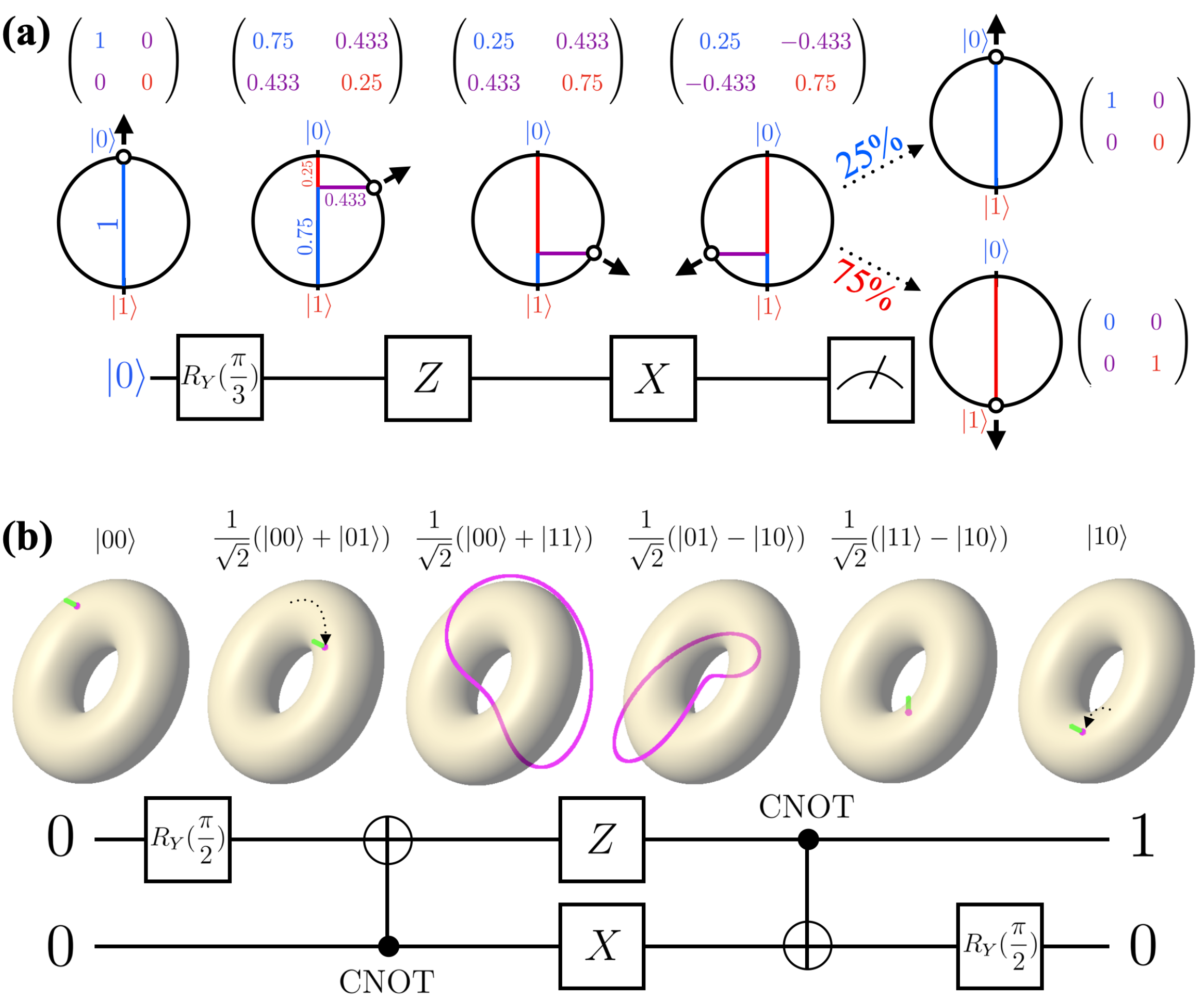}
\caption{(a) Single-qubit state evolution as the qubit is sent through a series of quantum logic gates ending with a measurement.  The unit-diameter Bloch Circle visualization of the (white) statepoint evolution contains lines whose lengths allow the corresponding density matrix elements of each state, which include measurement probabilities, to be read off directly.  The $R_Y$ gate rotates the statepoint on the circle, the $Z$ gate flips the statepoint about circle's horizontal diameter and the $X$ gate flips the statepoint about the vertical diameter (b) Two-qubit state evolution through a series of logic gates.  The two-qubit statepoint (pink) at the end of the constituent qubit radius (green) map the states, evolving in a toroidal region that contains the shown separable torus.  The first and last $R_Y$ gates rotate the first and second qubits and correspond to the statepoint rotating on the separable torus surface in the poloidal and toroidal directions respectively.  The controlled not (CNOT) gates take the qubits to and from maximally entangled states which are mapped as distinct torus knots that wind around the surface of the toroidal space.  The $Z$-$X$ gate combination changes the two-qubit state from one maximally entangled state to another and is equivalent to rotating the torus knot about the toroid axis by $180^\circ$.}
\label{figLogicGates}
\end{figure*}

The simplest 2-qubit systems are the separable states, those with no entanglement that are composed of the tensor product of two pure 1-qubit states:
\begin{equation}
\begin{split}
\ket{\chi}&=\ket{\psi_1} \otimes \ket{\psi_2} = \ket{\psi_1 \psi_2} \\ &\rightarrow \begin{pmatrix}a_1 \\ b_1 \end{pmatrix} \otimes \begin{pmatrix}a_2 \\ b_2 \end{pmatrix} = \begin{pmatrix}a_1a_2 \\ a_1b_2 \\ b_1a_2 \\ b_1b_2 \end{pmatrix}=\begin{pmatrix}\alpha \\ \beta \\ \gamma \\ \delta \end{pmatrix}
\label{eq:twoqubit_pure}
\end{split}
\end{equation}
\begin{figure}[ht!]
\centering
\includegraphics[scale=.19]{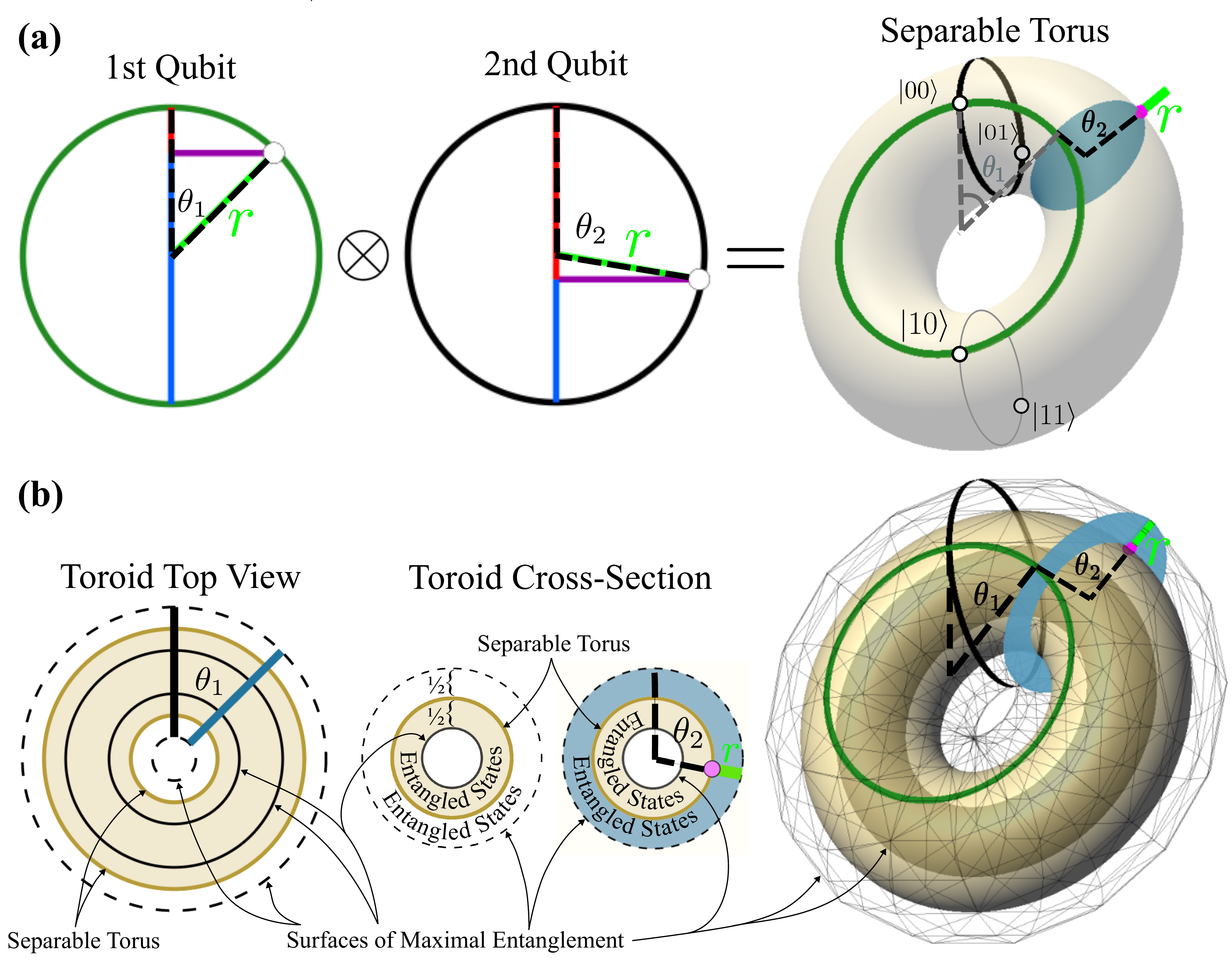}
\caption{(a) Two 1-qubit pure states, each plotted on Bloch Circles at angles $\theta_1$ and $\theta_2$, are encoded in a torus map of a separable 2-qubit state where the location of a (pink) statepoint at the end of the radius $r$ uniquely determines the two angles with respect to the (green and black) rings on the torus surface. Four (white) dots map the basis states $ \ket{00}, \ket{01}, \ket{10}, \text{and} \ket{11}$. (b) The separable torus surface in (a) is expanded outward and inward to form a toroidal volume with a cross-section in the shape of an annulus. The separable torus is embedded in the center of this toroid with the line from the surface of the toroid to the statepoint of length $r={1}/{2}$ corresponding to the radial coordinate of the individual 1-qubit states. Points in the toroid volume not on this embedded separable torus surface correspond to entangled qubit states (see also Fig.~\ref{figToroid}).}
\label{figSeparableTorus}
\end{figure}
As the general 1-qubit states are each independently parametrized by a polar and azimuthal angle corresponding to their Bloch Sphere statepoint location, a full accounting of only the separable 2-qubit states would require a 4-D space in which to plot a 2-qubit statepoint. We again circumvent the challenge of this visualization by focusing on a subspace of these states formed from tensor products of 1-qubit states located on the Bloch Circle. This restriction results in column vector representations with real entries, the first of which, $\alpha$, is always positive or zero. Since each of these 1-qubit states is parametrized by angles $\theta_1$ and $\theta_2$ ranging from $-\pi$ to $\pi$ on their respective Bloch Circles, e.g. $a_1=\cos\left(\theta_1/2\right)$, a natural visualization of the corresponding 2-qubit state is a statepoint on a surface that can encode each of these angles, namely a torus. Fig.~\ref{figSeparableTorus}(a) shows such a statepoint where the angles of the corresponding 1-qubit statepoints can be read off using the location of a separable 2-qubit statepoint on this ``separable torus.'' The statepoint locations of the standard 2-qubit orthonormal basis set, used to expand the general pure state $\ket{\chi}=\alpha\ket{00}+\beta\ket{01}+\gamma\ket{10}+\delta\ket{11}$, are labeled on the same figure for reference.  The evolution of the 2-qubit statepoint on the separable torus as the individual 1-qubit states are rotated can be seen in the first and last logic gates of Fig.~\ref{figLogicGates}(b).

To extend the visualization of statepoints to entangled (non-separable) pure 2-qubit states, we first note that the full space of these states can be parametrized by the four complex coefficients $\alpha,\beta,\gamma,\delta$ constrained by a bounding normalization condition $|\alpha|^2+|\beta|^2+|\gamma|^2+|\delta|^2=1$ and subjected to a choice of global phase that allows $\alpha$ to be taken to be real and non-negative. The resulting space is a 6-D projective plane, which involves both an unvisualizably high dimension as well as a non-trivial topology that is challenging to visualize even in lower dimensions. To circumvent these challenges, we follow our established approach of looking at only the real-coefficient subspace of the full state space, which, with the $\alpha\geq0$ phase choice and normalization condition, is a bounded 3-D region. The visualization of this subset of states and their corresponding Bloch Circle 1-qubit states is aided by the fact -- proved in Appendix \ref{proofSameRadius} -- that though the angles of the 1-qubit statepoints are independent, their degree of mixedness, as measured by the length of their radius $r$ or the length of the corresponding $s$ parameter, are the same for each qubit. Also shown in Appendix \ref{proofSameRadius} is that this degree of mixedness correlates with the degree of entanglement of the 2-qubit state and can be calculated directly from the 2-qubit expansion coefficients $s = {\alpha \delta - \beta\gamma}$, which has a magnitude $|s| = \sqrt{ \left({1}/{2}\right)^2 - r^2}$ that varies from 0 for separable states up to ${1}/{2}$ for maximally entangled states as each of the 1-qubit states vary from pure ($r={1}/{2}$) to maximally mixed ($r=0$).  While $r$ is a more directly intuited length for a statepoint on a Bloch Circle, the $s$ entanglement parameter magnitude also corresponds to a geometrical length (see Fig.~\ref{figMixedStates}(a)) and is simpler to calculate from the expansion coefficients in the 2-qubit case.  Most importantly, in what follows its sign can be used in our 2-qubit plotting scheme to encode an additional bit of information about which state in our subspace, with otherwise identical radii and sets of angles, is being specified.
\begin{figure}[h!]
\centering
\includegraphics[scale=.24]{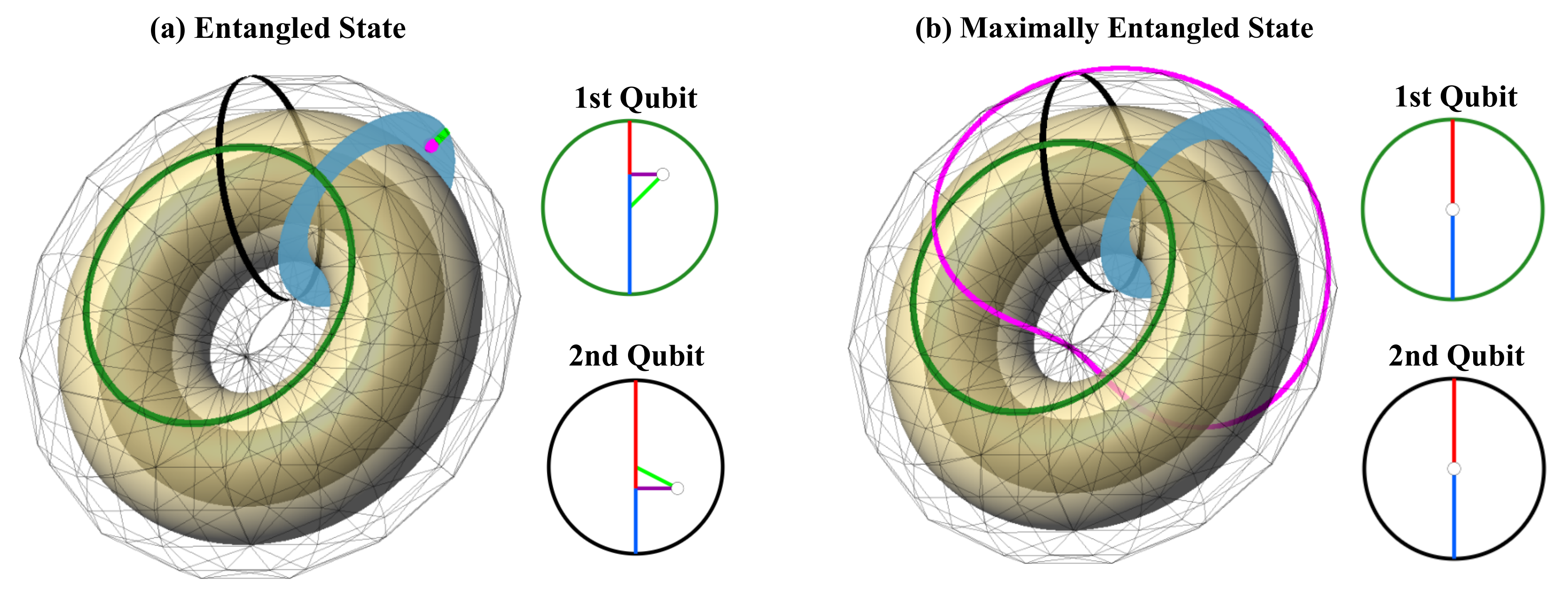}
\caption{Entangled states. (a) An entangled 2-qubit statepoint plotted on the 2-qubit toroid next to the corresponding 1-qubit states plotted on Bloch Circles. The pink statepoint, being off the embedded central separable torus, indicates it is an entangled state with the degree of entanglement parametrized by the green-line distance between the statepoint and the toroid surface. This length correlates to the shared radial length of each of the 1-qubit states and monotonically decreases with increasing entanglement. As in Fig.~\ref{figSeparableTorus}(a), the angular coordinates of the 1-qubit state can again be read off from the angular location of the statepoint in the toroid. (b) For each of the infinitely many maximally entangled 2-qubit states, the state is plotted not as a point but as a closed-curve torus knot on either the inner or outer surfaces of the toroid. The corresponding 1-qubit states are always maximally mixed with a 50/50 probability outcome for all possible measurements.}
\label{figToroid}
\end{figure}
To plot statepoints of these 2-qubit states in a bounded 3-D space, our three real, bounded parameters are thus the 1-qubit angles $\theta_1$, $\theta_2$, and the $s$ entanglement parameter. Starting from just the separable 2-qubit states, which have zero entanglement ($s=0$), and the pair of angles parameterized on the separable torus surface, we can extend this torus shell inward (for $s<0$) and outward (for $s>0$) by a ${1}/{2}$ magnitude to create a volume to map all the real 2-qubit states over the full range of $s$. The separable torus can be taken to have a cross-sectional radius of any size greater than ${1}/{2}$, so that the inward extension leaves an empty toroidal space inside of the toroidal volume rather than resulting in a filled-in ``donut.'' The resulting state space is thus an annular toroidal volume bounded by two torus surfaces (see Fig.~\ref{figSeparableTorus}(b) and Fig.~\ref{figMapping} in Appendix \ref{proofParameterization}). Any statepoint not on the central separable shell where $s=0$ ($r={1}/{2}$) is an entangled state (see the Fig.~\ref{figSeparableTorus}(b) cross-section), with states getting more entangled as they approach the outer ($s={1}/{2}$) and inner ($s=-{1}/{2}$) torus surfaces of this volume. As with the separable states, the angles $\theta_1$ and $\theta_2$ of the 1-qubit states can be read off directly from the location of the 2-qubit statepoint in this volume, with the shared radius $r$ of the 1-qubit states directly visualizable as the distance from the statepoint to the surface of the toroid as in Fig.~\ref{figToroid}(a). 

While 2-qubit states in the subspace typically have a one-to-one correspondence to points in the volume of the toroid, an exception is the class of maximally entangled states such as Bell states.  For these states, the 1-qubit states are maximally mixed, corresponding to statepoints at the center of the Bloch circles with no $\theta_1$ or $\theta_2$ angles and $r=0$ ($s=\pm{1/2}$).  Rather than being mapped to a point on the toroid, each of these states is mapped one-to-one to a unique closed-loop torus knot on either the inner or outer surface that winds once around the surface with respect to each angle (Fig.~\ref{figToroid}(b) and Appendix \ref{entangled}).  For example the state $\left({1}/{\sqrt{2}}\right)(\ket{01}+\ket{10})$ corresponds to a torus knot on the inner surface, while the state $\left({1}/{\sqrt{2}}\right)(\ket{01}-\ket{10})$ corresponds to a torus knot on the outer surface; the central logic gates in Fig.~\ref{figLogicGates}(b) transform one outer-surface torus knot state to another. The mapping of the maximally entangled states to knots rather than points is a manifestation of the non-trivial projective plane topology of the full state space.
The apparent drawback of mapping maximally entangled states to closed loops rather than points is compensated by the fact that this mapping helps to reveal one reason why these states play an important role in quantum information algorithms: they directly connect states with widely different 1-qubit angles.  For all statepoints in the toroid map, neighboring statepoints correspond to neighboring states in the Hilbert space, so that a point moving along a continuous path through the 2-qubit toroid volume corresponds to a unitary evolution of a 2-qubit state within the real-coefficient subspace. The fact that neighboring points correspond to neighboring states is preserved for the maximally entangled states as well -- all points in the geometric neighborhood of one these closed loops are also neighbors in the Hilbert space.  Thus each loop corresponding to a maximally entangled state acts as a ``bridge'' one can evolve through in a continuous fashion in order connect statepoints in different sections of the toroid:  a statepoint near the toroid surface can evolve into a state mapped by a loop and then evolve into any statepoint near this loop in any other part of the toroid.  As proved in Appendix \ref{proofParameterization}, there is an outer-surface knot connecting states for each unique fixed sum of 1-qubit angles $\theta_1+\theta_2$ and an inner-surface knot connecting states for each unique fixed difference of 1-qubit angles $\theta_1-\theta_2$.

Finally, one intuition about the correlations between 2-qubit pure states and their corresponding 1-qubit states that can be reinforced by these visualizations is that there is not a simple one-to-one mapping between them. While the separable 2-qubit states map to unique 1-qubit states in this subspace (reflective of a one-to-one mapping in the full space up to a relative phase between the individual qubits), there is a two-to-one mapping for the general entangled 2-qubit states (one for each possible sign of the same-magnitude $s$ parameter at the same $\theta_1$ and $\theta_2$ angles) and infinitely many maximally entangled states corresponding to the 1-qubit pair where each qubit is in its maximally mixed state (reflective of the continuum of torus knots on the outer and inner surfaces of the toroid).

\section{Conclusion}
The increasing need to introduce basic quantum information concepts at multiple education levels, including teaching students with formal backgrounds and future prospects not limited to conventional physics training, necessitates the development of new tools for conveying these concepts that do not rely on first mastering the linear algebra quantum formalism. A virtue of the subspace visualization approach presented here is that it can be deployed as a pedagogical tool in different ways for different audiences. For students with minimal mathematical training, the points and lines of the Bloch Circle develop intuitions about qubit states, pure and mixed; choice of measurement and bases; and probabilities and probability amplitudes; while the 2-qubit toroid develops intuitions about the difference between separable and entangled states; the relation of pure 2-qubit states to the 1-qubit states of their entangled qubits; and the multiple paths of unitary evolutions of these states. For students with some background in linear algebra, these visualizations can aid in understanding the actions of algorithms in the quantum formalism that can be restricted to real-number calculations. Indeed, in addition to the measurement algorithm, 1-qubit unitary quantum logic gates such as R$_{\text{Y}}$, X, Z, and Hadamard, and 2-qubit gates such as CNOT, Controlled Z, and SWAP are closed transformations within the real-state subspace whose actions can be visualized, such as in the examples shown in Fig.~\ref{figLogicGates}. The toroid map means that single-qubit rotations in the real subspace correspond to rotations of the statepoint or knot about the toroidal or poloidal directions, depending on which qubit is being rotated.  The evolution of the statepoint due to other gates is less immediately intuitive -- like all maps the correspondence between physical state evolution and visual representation is built through practice and specific examples. The proofs in the Appendices relating calculations in the quantum formalism to lengths and angles in the diagrams, can form the basis for multiple exercises at the more advanced undergraduate level, where visual intuition and mastery of the formalism can be developed in tandem. 
\appendix

\section{Geometric Interpretation of the Bloch Circle} \label{proofGeoProp}
In this appendix we prove the equivalence between geometrical lengths drawn on a Bloch Circle and the magnitudes of matrix elements and measurement probabilities for 1-qubit state representations.

\begin{figure}[h!]
\centering
\includegraphics[scale=.17]{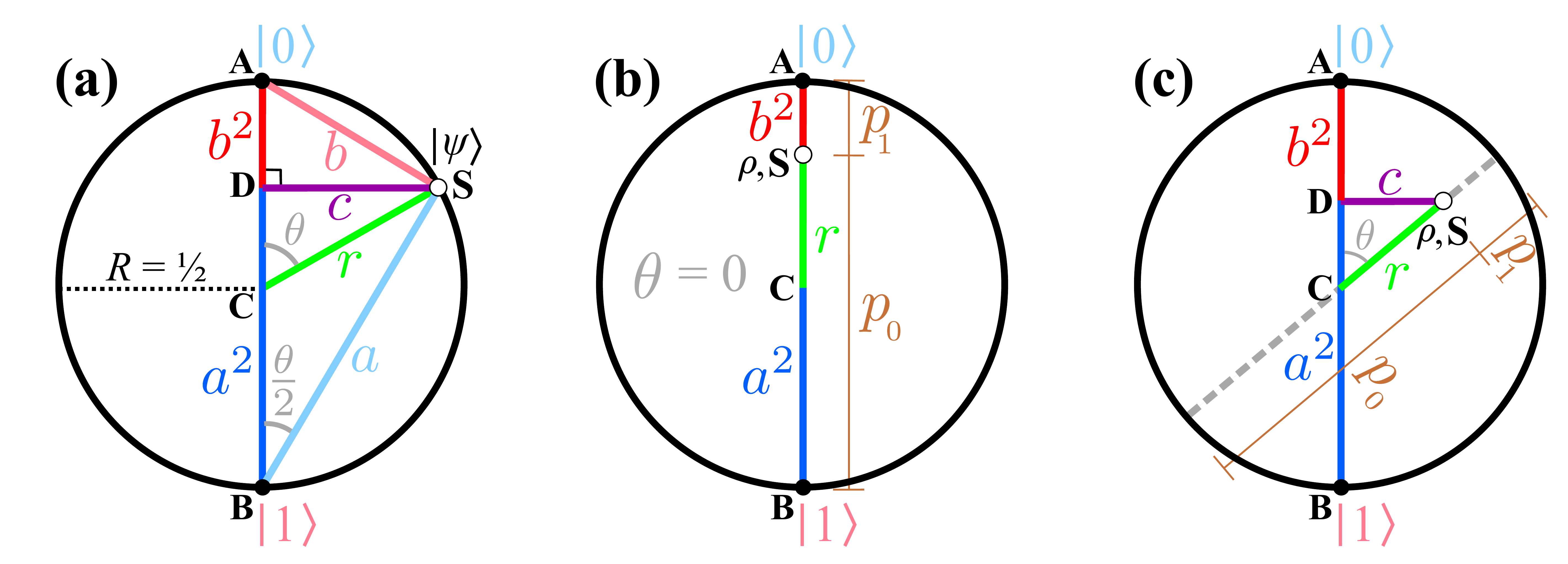}
\caption{ (a) A pure 1-qubit state $\ket{\psi}$ that lies on the periphery of the Bloch Circle with line segments formed as in Fig.~\ref{figBlochCircle}. (b) A statepoint $S$ plotted along the vertical basis axis corresponding to the diagonalized density matrix of Eq.~\ref{eq:p0_p1_matrix}. (c) The statepoint for the mixed state in (b) when the 1-qubit state has been rotated by an angle $\theta$.  Geometrical lengths formed by drawing a perpendicular to the basis axis correspond to elements of the state's density matrix in Eq.~\ref{eq:rho_color2}.}
\label{figBlochGeometry}
\end{figure}

In Fig.~\ref{figBlochGeometry}(a) a point $S$ is plotted on a Bloch Circle with radius $R={1}/{2}$ to represent a pure state $\ket{\psi}$. A basis of representation/measurement is chosen by drawing any diameter of the circle $\overline{AB}$ and designating the points $A$ and $B$ to be the states
\begin{equation}
\ket{0} \rightarrow \begin{pmatrix} 1 \\ 0 \end{pmatrix} \quad \text{and} \quad \ket{1} \rightarrow \begin{pmatrix} 0 \\ 1 \end{pmatrix},
\label{eq:0_1_vectors}
\end{equation}
 respectively. Lines are drawn between these three points and the center $C$ of the Bloch Circle, and a line from $S$ perpendicular to $\overline{AB}$ is drawn that intersects this diameter at point $D$.
The location of the pure statepoint $S$ relative to the chosen basis can be parametrized either by the angle $\theta=\angle{ACS}$ at the center of the circle or $\theta/2=\angle{ABS}$, which is half of $\angle{ACS}$ as it subtends the same arc as this angle but is drawn from the circle circumference\cite{euclid1}.  Since $\angle{ASB}$ subtends the diameter of the circle it is a right angle\cite{euclid2}, meaning $\triangle ABS$ is a right triangle with unit hypotenuse and $\theta/2$ acute angle.  Thus we can alternately parametrize the state $S$ in this basis by specifying the lengths of the legs of this triangle,
\begin{equation}
\ket{\psi} \rightarrow \begin{pmatrix} a \\ b \end{pmatrix} \equiv \begin{pmatrix} \overline{BS} \\ \overline{AS} \end{pmatrix}=\begin{pmatrix} \cos(\tfrac{\theta}{2}) \\ \sin(\tfrac{\theta}{2}) \end{pmatrix},
\label{eq:purestate_column}
\end{equation}
which is the conventional column-vector representation of a pure qubit state in the real-state subspace corresponding to a cross-section of the Bloch Sphere taken at azimuthal angle $\phi=0$. Note the convention that $\overline{AS}=b$ is positive for statepoints on the right half of the circle, negative on the left half, and that it would be multiplied by $e^{i\phi}$ if other azimuthal angle cross-sections were chosen.

To visualize the density matrix representation of this same state in this basis, we note that from the similar right triangles $\triangle ASB \sim \triangle ADS \sim \triangle SDB$ we have the equal ratios ${\overline{BD}}/{\overline{BS}}={\overline{BS}}/{\overline{AB}}\Rightarrow\overline{BD}=a^2$, ${\overline{AD}}/{\overline{AS}}={\overline{AS}}/{\overline{AB}}\Rightarrow\overline{AD}=b^2$, and ${\overline{SD}}/{\overline{AS}}={\overline{BS}}/{\overline{AB}}\Rightarrow\overline{SD}=ab$. These three line segments are thus equal in magnitude to the density matrix elements:
\begin{equation}
\ket{\psi} \bra{\psi} \rightarrow \rho = \begin{pmatrix} a^2 & ab \\ ab & b^2 \end{pmatrix} = \begin{pmatrix} \overline{BD} & \overline{SD} \\ \overline{SD} & \overline{AD} \end{pmatrix}.
\label{eq:psi_bra_psi_rho}
\end{equation}
Note again the convention that the off-diagonal element $\overline{SD}=ab\equiv{c}$ can be taken to be positive for statepoints on the right half of the circle and modified by $e^{\pm{i\phi}}$ for other subspace cross-sections. The diagonal matrix elements are always non-negative, with the two line segments $\overline{BD}$ and $\overline{AD}$ of the chosen measurement diameter exactly equal to the probabilities that the state $S$ will become the state at $A$ or $B$ on measurement.

The cutting of the basis diameter into pieces equal to probabilities of measurement outcomes can be extended to plotting mixed statepoints. Consider a state that results from a qubit having the epistemic probabilities $p_0$ and $p_1$ of being in the states $\ket{0}$ or $\ket{1}$ respectively. This state is represented by the diagonalized density matrix
\begin{equation}
p_0 \begin{pmatrix} 1 \\ 0 \end{pmatrix}\begin{pmatrix} 1 & 0 \end{pmatrix} + p_1 \begin{pmatrix} 0 \\ 1 \end{pmatrix}\begin{pmatrix} 0 & 1 \end{pmatrix} = \begin{pmatrix} p_0 & 0 \\ 0 & p_1 \end{pmatrix}
\label{eq:p0_p1_matrix}
\end{equation}
which can be plotted as a point along the measurement diameter that cuts it into segments equal to these probabilities as shown in Fig.~\ref{figBlochGeometry}(b).

To create a general mixed-state point within the circle, we note that the probabilities $p_0$ and $p_1$ determine the radial coordinate of the statepoint, $r = (p_0 - p_1)/2$, and that changing the angular coordinate of the statepoint can correspond to a rotational transformation on the density matrix by
\begin{equation}
R(\theta) \equiv \begin{pmatrix} \cos(\tfrac{\theta}{2}) & -\sin(\tfrac{\theta}{2}) \\ \sin(\tfrac{\theta}{2}) & \cos(\tfrac{\theta}{2})
\end{pmatrix}.
\label{eq:rotation_matrix}
\end{equation}
Thus the state 
\begin{equation}
\begin{split}
\rho &= R(\theta) \begin{pmatrix} p_0 & 0 \\ 0 & p_1 \end{pmatrix} R^\dagger(\theta) \\&= \begin{pmatrix} p_0 \cos^2(\tfrac{\theta}{2}) + p_1\sin^2(\tfrac{\theta}{2}) & (p_0-p_1) \sin(\tfrac{\theta}{2})\cos(\tfrac{\theta}{2}) \\ (p_0-p_1) \sin(\tfrac{\theta}{2})\cos(\tfrac{\theta}{2}) & p_1 \cos^2(\tfrac{\theta}{2}) + p_0\sin^2(\tfrac{\theta}{2}) \end{pmatrix}
\label{eq:rho_rotated}
\end{split}
\end{equation}
corresponds to the statepoint shown in \ref{figBlochGeometry}(c). Again drawing a perpendicular from this point to cut the measurement axis $\overline{AB}$ at point $D$, we have the following geometric relations: 
\begin{equation}
\begin{split}
\overline{SD} &= r\sin(\tfrac{\theta}{2} + \tfrac{\theta}{2}) = \tfrac{1}{2} (p_0-p_1) \times 2 \cos{(\tfrac{\theta}{2})} \sin{(\tfrac{\theta}{2})} 
\\ &= (p_0-p_1) \sin(\tfrac{\theta}{2}) \cos(\tfrac{\theta}{2}), 
\\ \overline{AD} &= R - \overline{DC} = \tfrac{1}{2} -r\cos(\tfrac{\theta}{2} + \tfrac{\theta}{2}) 
\\ &=
	\tfrac{1}{2}(p_0+p_1)\left(\cos^2{(\tfrac{\theta}{2})} + \sin^2{(\tfrac{\theta}{2}})\right)
\\	&\quad\quad-\tfrac{1}{2}(p_0-p_1)\left(\cos^2{(\tfrac{\theta}{2})} - \sin^2{(\tfrac{\theta}{2})}\right) 
\\ &=p_1 \cos^2(\tfrac{\theta}{2}) + p_0\sin^2(\tfrac{\theta}{2}),
\\ \overline{BD} &= 1 - \overline{AD}
\\ &=\left(p_0+p_1\right)\left(\cos^2{(\tfrac{\theta}{2})} +\sin^2{(\tfrac{\theta}{2})}\right)
\\ &\quad\quad-\left(p_1 \cos^2(\tfrac{\theta}{2}) + p_0\sin^2(\tfrac{\theta}{2})\right) 
\\ &=p_0\cos^2{(\tfrac{\theta}{2})} + p_1\sin^2{(\tfrac{\theta}{2})}.
\label{eq:lengths}
\end{split}
\end{equation}
By matching matrix elements of Eq.~\ref{eq:rho_rotated} with these line segments we have the geometric interpretation of a Bloch Circle for a general mixed state where the diagonal line segments again correspond to measurement probabilities for the chosen basis diameter and the magnitude of the off-diagonal element is equal to the distance from the statepoint to this diameter:
\begin{equation}
\rho = \begin{pmatrix} \overline{BD} & \overline{SD} \\ \overline{SD} & \overline{AD} \end{pmatrix}
\label{eq:rho_geometric}
\end{equation}

\section{Shared Radius of Constituent Qubits of 2-qubit States} \label{proofSameRadius}
A pure 2-qubit state $\ket{\chi}$ in the real-state subspace is represented by a column vector with real entries:
\begin{equation}
\ket{\chi} \rightarrow \begin{pmatrix} \alpha \\ \beta \\ \gamma \\ \delta \end{pmatrix}
\label{eq:chi_vector}
\end{equation}

The normalization condition $\alpha^2+\beta^2+\gamma^2+\delta^2=1$ bounds the magnitude of the entries, and, with the choice of a global phase, we always take $\alpha$ to be positive when non-zero, so that $0\leq\alpha\leq1$, and $ -1 \leq \beta , \gamma ,\delta \leq 1$.
A representation of the density matrix $\rho$ of this state is obtained by taking the outer product of the column vector representation:
\begin{equation}
\ket{\chi} \bra{\chi} \rightarrow
\begin{pmatrix} \alpha \\ \beta \\ \gamma \\ \delta \end{pmatrix}
\begin{pmatrix} \alpha & \beta & \gamma & \delta \end{pmatrix} = \rho = \begin{pmatrix} 
\alpha^2 & \alpha \beta & \alpha \gamma & \alpha \delta \\
\alpha \beta & \beta^2 & \beta \gamma & \beta \delta \\
\alpha \gamma & \gamma \beta & \gamma^2 & \gamma \delta \\
\alpha \delta& \beta\delta & \gamma \delta & \delta^2 \\
\end{pmatrix}.
\label{eq:chi_rho_matrix}
\end{equation}
Density matrices for the constituent single qubits are found by taking two partial traces of $\rho$:
\begin{equation}
\begin{split}
\rho_1 &= Tr(\rho,1) = \begin{pmatrix} 
\alpha^2+\gamma^2 && \alpha \beta+ \gamma\delta \\
\alpha \beta+ \gamma\delta && \beta^2+\delta^2 
\end{pmatrix} 
\equiv \begin{pmatrix} x && y \\ y && z \end{pmatrix}
\\
\rho_2 &= Tr(\rho,2) = \begin{pmatrix} 
\alpha^2+\beta^2 && \alpha \gamma+\beta\delta \\ 
\alpha \gamma+\beta\delta && \gamma^2+\delta^2 \end{pmatrix}
\equiv \begin{pmatrix} w && u \\ u && v \end{pmatrix}.
\end{split}
\label{eq:partial_traces}
\end{equation}
Each of these states can be plotted as statepoints on a Bloch Circle, located at distances $r_1$ and $r_2$ from the center respectively. From the geometric properties derived in Appendix \ref{proofGeoProp} these distances are the hypotenuses of right triangles satisfying
\begin{equation}
\begin{split}
r_1^2 &= (x-\tfrac{1}{2})^2 + y^2 = (\alpha^2+\gamma^2-\tfrac{1}{2})^2 + (\alpha \beta+ \gamma\delta)^2, \\
r_2^2 &= (w-\tfrac{1}{2})^2 + u^2 = (\alpha^2+\beta^2-\tfrac{1}{2})^2 + (\alpha \gamma+\beta\delta)^2.
\end{split}
\label{eq:r_alternate}
\end{equation}
Subtracting these two equations, regarding the difference of the middle terms in Eq.~\ref{eq:r_alternate} as the difference of squares, using the definitions of Eq.~\ref{eq:partial_traces}, and recalling the normalization condition, we find
\begin{equation}
\begin{split}
r_1^2 - r_2^2 
&= \left[\left(x-\tfrac{1}{2}\right)+\left(w-\tfrac{1}{2}\right)\right]\left[\left(x-\tfrac{1}{2}\right) -\left(w-\tfrac{1}{2}\right)\right] \\
&\quad\quad+ \left(y^2-u^2\right) \\
&= (x+w-1)(x-w) + \left(y^2-u^2\right) \\
&= \left[ \alpha^2 + (\alpha^2 + \gamma^2+\beta^2 -1)\right](\gamma^2-\beta^2)\\ 
&\quad\quad+ \left[(\alpha \beta)^2 +(\gamma \delta)^2-(\alpha \gamma)^2-(\beta\delta)^2\right] \\
&= (\alpha^2-\delta^2)(\gamma^2-\beta^ 2)+ (\alpha^2-\delta^2)(\beta^2-\gamma^2) =0.
\end{split}
\end{equation}
Since $r_1$ and $r_2$ are non-negative, it follows that $r_1=r_2\equiv{r}$. Thus the statepoints of the constituent single qubits of a pure 2-qubit state have the same radius $r$ on the Bloch Circle, a result which also holds for states in the full, pure 2-qubit state space and corresponding 1-qubit states plotted on a Bloch Sphere.

The shared $r$ is closely related to an alternate shared parameter that is useful for visualizing 2-qubit states: $s \equiv \alpha \delta - \beta \gamma$. The parameter $s$ is proportional to the quantum ``concurrence''\cite{concurrenceOriginal, concurrenceRev} of the qubit pair and is a direct measure of the degree of entanglement, varying from $s=0$ for separable states to $|s|={1}/{2}$ for maximally entangled states. This property may be seen in its relation to $r$, which can be calculated from the normalization condition and Eq.~\ref{eq:r_alternate}:
\begin{equation}
\begin{split}
s^2 &= \left(\alpha \delta - \beta\gamma\right)^2 \\
&= \tfrac{1}{4} - \tfrac{1}{4} + \left(\alpha^2 \gamma^2 + \alpha^2 \delta^2 + \beta^2 \gamma^2 + \beta^2 \delta^2\right)- \left(\alpha \gamma + \beta\delta\right)^2 \\
&= \tfrac{1}{4} - \bigg[\tfrac{1}{4}\bigg(\left(\alpha^2+\beta^2 + \gamma^2 + \delta^2\right)^2\\
&\quad\quad-4\left(\alpha^2 \gamma^2 + \alpha^2 \delta^2 + \beta^2 \gamma^2 + \beta^2 \delta^2\right)\bigg) + \left(\alpha \gamma + \beta\delta\right)^2\bigg] \\
&= \left(\tfrac{1}{2}\right)^2 -\left[\left(\tfrac{1}{2}\left(\alpha^2+\beta^2 - \gamma^2 - \delta^2\right)\right)^2 + \left(\alpha \gamma + \beta\delta\right)^2\right] \\
&= \left(\tfrac{1}{2}\right)^2 -\left[\left(\alpha^2+\beta^2 - \tfrac{1}{2}\right)^2 + \left(\alpha \gamma + \beta\delta\right)^2\right] \\ 
&= \left(\tfrac{1}{2}\right)^2 - r^2, \\ 
&\Rightarrow s = \pm \sqrt{\left(\tfrac{1}{2}\right)^2 - r^2}.
\end{split}
\label{eq:s-r_relation}
\end{equation}
Visually, this result shows that $|s|$ and $r$ form the legs of a right triangle with ${1}/{2}$ being the hypotenuse. From the geometry of the Bloch Circle (see Appendix \ref{proofGeoProp}), this means that the magnitude of $s$ is the length of the line segment drawn perpendicular to $r$ from the statepoint to the perimeter of the Bloch Circle (see Fig.~\ref{figMixedStates}(a)). The sign associated with $s$ allows the encoding of an additional bit of information useful in mapping the 2-qubit states. 
\section{Maximally Entangled States}
\label{entangled}
The results of Appendix \ref{proofSameRadius} allow us to write explicit column-vector representations of all maximally entangled 2-qubit states. The $|s|={1}/{2}$ or $r=0$ condition entails that the 1-qubit density matrices have the form
\begin{equation}
\rho_1 = \rho_2 = \begin{pmatrix} \frac{1}{2} & 0 \\ 0 & \frac{1}{2} \end{pmatrix}
\label{eq:rho_1_rho_2_matrix}
\end{equation}
and are plotted at the center of the Bloch Circles. From the geometry of Eq.~\ref{eq:psi_bra_psi_rho} and the diagonal elements of Eq.~\ref{eq:partial_traces}, the column vector entries of the maximally entangled 2-qubit states must satisfy $\alpha^2+\gamma^2=\beta^2+\delta^2=\alpha^2+\beta^2=\gamma^2+\delta^2={1}/{2}$, so that each coefficient has a maximum magnitude of ${1}/{\sqrt{2}}$. Again using our choice of global phase to always take $\alpha$ to be positive when non-zero, we can parametrize all its possible values for a maximally entangled state using a single angle $\xi$ ranging from $-\pi$ to $\pi$, $\alpha=\left({1}/{\sqrt{2}}\right)\cos{\left({\xi}/{2}\right)}$, and use the diagonal element equalities to write the other possible coefficient values as $\beta=\pm\left({1}/{\sqrt{2}}\right)\sin{\left({\xi}/{2}\right)}$, $\gamma=\pm\left({1}/{\sqrt{2}}\right)\sin{\left({\xi}/{2}\right)}$, and $\delta=\pm\left({1}/{\sqrt{2}}\right)\cos{\left({\xi}/{2}\right)}$. The relative signs of these coefficients are fixed by the off-diagonal equalities of the 1-qubit density matrices, $\alpha\beta+\gamma\delta=\alpha\gamma+\beta\delta=0$, which dictates the two possible sets of maximally entangled states

\begin{equation}
\ket{\chi_+} \rightarrow \frac{1}{\sqrt2}\begin{pmatrix}
\cos( \frac{\xi}{2} ) 
\\ -\sin( \frac{\xi}{2} ) 
\\ \sin( \frac{\xi}{2} ) 
\\ \cos( \frac{\xi}{2} ) 
\end{pmatrix}\;\textrm{and}\;
\ket{\chi_-}
\rightarrow \frac{1}{\sqrt2} \begin{pmatrix} 
\cos( \frac{\xi}{2} ) 
\\ \sin( \frac{\xi}{2} ) 
\\ \sin( \frac{\xi}{2} ) 
\\ -\cos( \frac{\xi}{2} ) 
\end{pmatrix}.
\label{eq:chis}
\end{equation}
The subscripts labeling these sets are chosen to reflect the fact that $s=+{1}/{2}$ for the first set and $s=-{1}/{2}$ for the second. All possible maximally entangled 2-qubit states in the real-state subspace are thus specified up to a global phase by this choice of sign and a single angle over a $2\pi$ range.

\section{Mapping 2-Qubit States} \label{proofParameterization}
In this section we establish formulas for the mappings between the 1-qubit geometric parameters $s, \theta_1, \theta_2$; the 2-qubit pure-state column vector entries $\alpha, \beta, \gamma, \delta$; and the statepoints on the two 1-qubit Bloch circles and 2-qubit annular toroid.
\begin{figure}[h!]
\centering
\includegraphics[scale=.55]{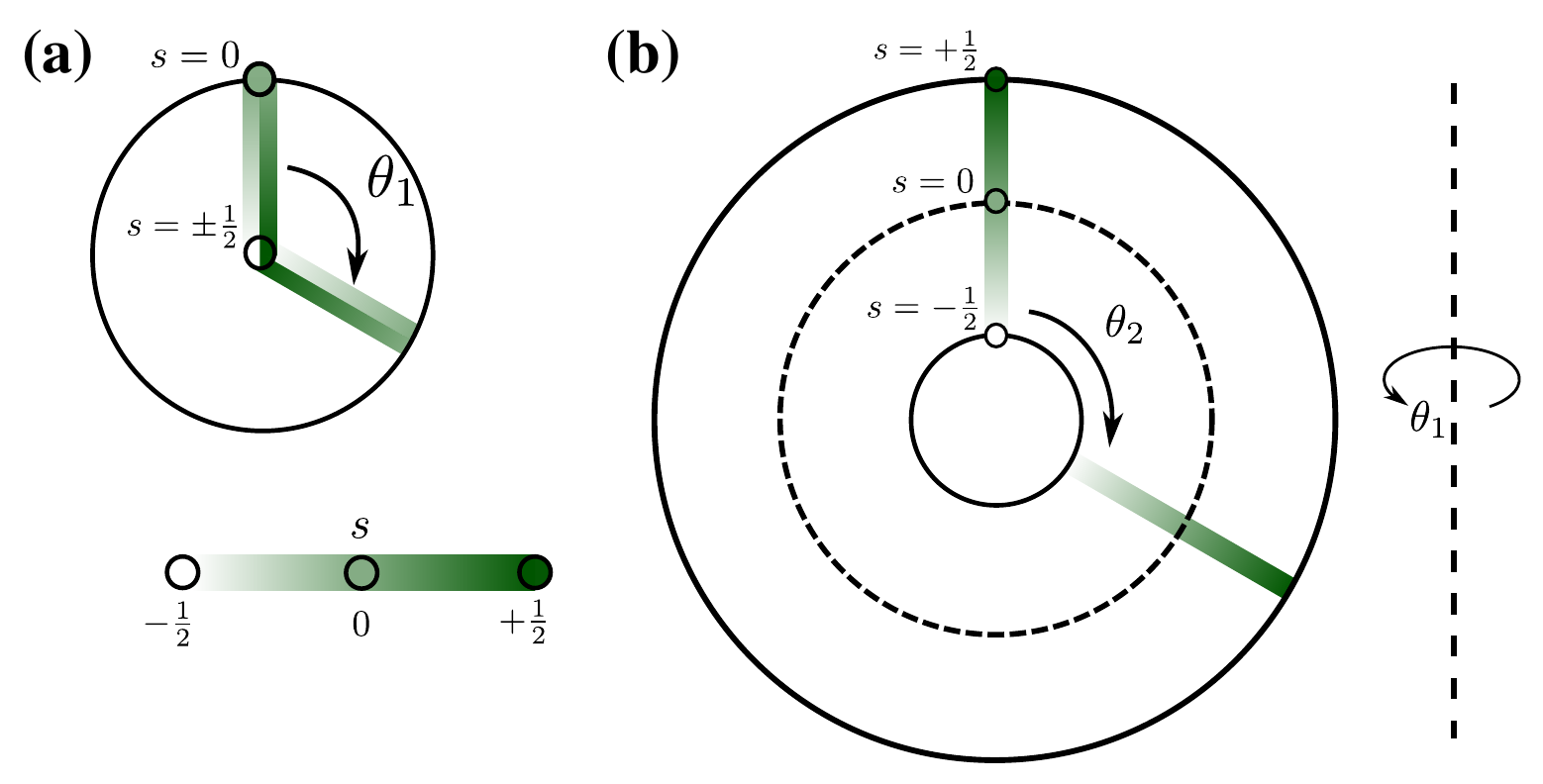}
\caption{(a) A gradient shows the possible $s$ values of pure 2-qubit quantum states as they range from $-{1}/{2}$ to ${1}/{2}$. As the shared radius of either 1-qubit statepoint $r=|s|$ evolves from ${1}/{2}$ to $0$ and back over this range, a statepoint corresponding to states with $\theta=0$ traces out a vertical radius twice over this range as it moves from the center of the Bloch Circle to its edge and back again. Statepoints corresponding to other angles can be graphically found by rotating this radius by the angle $\theta$ for each qubit. (b) To create a one-to-one correspondence between the 2-qubit geometric parameters $s,\theta_1,\theta_2$ and statepoints, the $s$ values are mapped one-to-one to a vertical line. With fixed $\theta_1$, all possible $\theta_2$ statepoints are created by rotating this line about a collinear center point below it by $\theta_2$ to form an annulus. All possible $\theta_1$ statepoints are created by rotating this annulus about a vertical axis outside of it by $\theta_1$ to form a toroid.} 
\label{figMapping}
\end{figure}
Starting with the mapping between the geometric parameters and statepoints, consider the plotting of states for all possible $s$ values for the case $\theta_1= \theta_2=0$. As shown in Appendix \ref{proofSameRadius} and Fig.~\ref{figMapping}(a), while $s$ ranges from $-{1}/{2}$ to ${1}/{2}$ either of the 1-qubit statepoints will move from the center of the Bloch Circle to the edge and back again along the vertical radius corresponding to $\theta=0$. Each statepoint along this radius thus corresponds to two values of $s$ with the exception of the pure state $s=0$ statepoint at the edge. 1-qubit statepoints for other angles from $-\pi$ to $\pi$ for each qubit are found by rotating this vertical radius to sweep out the full circle. The general two-to-one mapping of $s, \theta_1, \theta_2$ parameters and statepoints holds with the exception of the pure-state perimeter, where the mapping is one-to-one, and the maximally mixed point at the center, where there infinitely many choices of $\theta_1, \theta_2$ that map to this statepoint for the single qubits.

To make the geometric mapping everywhere one-to-one, we first map the $s$ values when $\theta_1=\theta_2=0$ to statepoints on a vertical line in a one-to-one fashion as shown in Fig.~\ref{figMapping}(b). Other values of $\theta_2$ are plotted by rotating this line about a central point below it to form an annulus, and other values of $\theta_1$ are plotted by rotating this annulus azimuthally about a vertical axis outside of it into the 3rd dimension to form the annular toroid shown in Fig.~\ref{figSeparableTorus}.

To correlate the 2-qubit column-vector entries to this mapping, consider the following states $\ket{\chi_0}$ that map one-to-one to all possible values of $s$:
\begin{equation}
\begin{split}
\ket{\chi_0}&\rightarrow\begin{pmatrix} \sqrt{\frac{1}{2} + \sqrt{ \left(\frac{1}{2}\right)^2 - s^2 } } \\ 0\\ 0\\ \text{sgn} (s) \sqrt{\frac{1}{2} - \sqrt{\left(\frac{1}{2}\right)^2- s^2 }} 
\end{pmatrix} \\
&=
\begin{pmatrix} \sqrt{\frac{1}{2} + r} \\ 0\\ 0\\ \\ \text{sgn} (s)\sqrt{\frac{1}{2} - r} 
\end{pmatrix}
\equiv \begin{pmatrix} \alpha_0 \\ 0\\ 0\\ \delta_0
\end{pmatrix}.
\end{split}
\end{equation}
The normalized coefficients $\alpha_0\geq\delta_0$ vary continuously over the full range of $s$ producing states in the chosen real-state subspace. From Eq.~\ref{eq:partial_traces}, the density matrices of the single qubits corresponding to $\ket{\chi_0}$ are
\begin{equation}
\rho_1 = \rho_2 = \begin{pmatrix} {\alpha_0}^2 && 0 \\ 0 && {\delta_0}^2 \end{pmatrix}.
\end{equation}
With off-diagonal elements equal to $0$ and ${\alpha_0}^2\geq{\delta_0}^2$, these states correspond to the continuum of statepoints lying along the $\theta_1=\theta_2=0$ vertical lines of Fig.~\ref{figMapping} for all possible $s$ values. To find 2-qubit column vectors for all other angles, we again sweep out this line by rotating each of the 1-qubit states. Formally, this is accomplished in the 2-qubit space using the tensor product of two of the 1-qubit rotation matrices $R(\theta)$ of Eq.~\ref{eq:rotation_matrix} to create a unitary rotation operator $R_{1,2}\equiv{R(\theta_1)\otimes{R(\theta_2)}}$ in the 2-qubit space and use this on $\ket{\chi_0}$:

\begin{equation}
\begin{split}
\ket{\chi} &\rightarrow R_{1,2}\begin{pmatrix} \alpha_0 \\ 0\\ 0\\ \delta_0 
\end{pmatrix}\\
&=
\begin{pmatrix} 
\alpha_0 \cos(\theta_1/2) \cos(\theta_2/2) + \delta_0 \sin(\theta_1/2) \sin(\theta_2/2) 
\\ \alpha_0 \cos(\theta_1/2) \sin(\theta_2/2) - \delta_0 \sin(\theta_1/2) \cos(\theta_2/2) 
\\ \alpha_0 \sin(\theta_1/2) \cos(\theta_2/2) - \delta_0 \cos(\theta_1/2) \sin(\theta_2/2) 
\\ \alpha_0 \sin(\theta_1/2) \sin(\theta_2/2) + \delta_0 \cos(\theta_1/2) \cos(\theta_2/2) 
\end{pmatrix}.\\
\end{split}
\end{equation}

This parametrization of $\ket{\chi}$ serves as a functional map from any choice of $s, \theta_1, \theta_2$ or their corresponding statepoints to the chosen real subspace of 2-qubit states. However, this map is not bijective, as can be seen by examining the maximally mixed $s=\pm{1}/{2}$ states that map to the outer and inner surfaces of the toroid. For the outer surface, $s={1}/{2},\alpha_0 = \delta_0 = {1}/{\sqrt2}$, and, with aid of trigonometric identities, the above mapping simplifies to

\begin{equation}
\ket{\chi_+} \rightarrow \frac{1}{\sqrt2} \begin{pmatrix} 
\cos( \frac{\theta_1 - \theta_2}{2} ) 
\\ -\sin( \frac{\theta_1 - \theta_2}{2} ) 
\\ \sin( \frac{\theta_1 - \theta_2}{2} ) 
\\ \cos( \frac{\theta_1 - \theta_2}{2} ) 
\end{pmatrix}.
\end{equation}
For the inner surface, $s=-{1}/{2}, \alpha_0 =-\delta_0 ={1}/{\sqrt2}$, and 
\begin{equation}
\ket{\chi_-}
\rightarrow \frac{1}{\sqrt2} \begin{pmatrix} 
\cos( \frac{\theta_1 + \theta_2}{2} ) 
\\ \sin( \frac{\theta_1 + \theta_2}{2} ) 
\\ \sin( \frac{\theta_1 + \theta_2}{2} ) 
\\ -\cos( \frac{\theta_1 + \theta_2}{2} ) 
\end{pmatrix}.
\end{equation}
Comparing these states with the full set of maximally entangled states in Eq.~\ref{eq:chis}, we see that the sign and angle needed to uniquely determine these states are given by $\text{\text{sgn} } (s=\pm{{1}/{2}})$ and $\xi=\theta_1\mp\theta_2$. With the angles restricted to a $2\pi$ range there will be infinitely many combinations of $\theta_1$ and $\theta_2$ that map to the same maximally mixed state. Geometrically the locus of all statepoints on either the outer or inner surface of the toroid ($\ket{\chi_+}$ or $\ket{\chi_-}$ states) corresponding at a single state with fixed $\xi=\theta_1\mp\theta_2$ value is a torus knot -- a closed loop as shown in Fig.~\ref{figToroid}(b) that winds once around each torus angle. The knots on the outer and inner surfaces wind with opposite helicity dependent on whether $\xi$ is determined by the difference or sum of $\theta_1$ and $\theta_2$.

Though this analysis shows there is a many-to-one mapping of the $s, \theta_1, \theta_2$ geometric parameters to unique maximally mixed states, there is a one-to-one mapping from these parameters to all other 2-qubit states in our real subspace as can be shown by establishing a function that maps $\alpha, \beta, \gamma, \delta$ to the geometric parameters.

Applying the geometric properties of the Bloch sphere derived in Appendix \ref{proofGeoProp}, the Eq.~\ref{eq:partial_traces} density matrices of the constituent qubits, and the relations of Eq.~\ref{eq:r_alternate} and Eq.~\ref{eq:s-r_relation}, we can find the geometric parameters $s, \theta_1, \theta_2$ in terms of the $u,w,x,y$ 1-qubit density matrix elements, which are in turn functions of the $\ket{\chi}$ vector coefficients $\alpha$, $\beta$, $\gamma$, and $\delta$:
\begin{equation}
\begin{split}
s &= \alpha \delta - \beta \gamma, \\
r &= \sqrt{ \left(\tfrac{1}{2}\right)^2 -s^2}=\sqrt{\left(x-\tfrac{1}{2}\right)^2 + y^2} = \sqrt{\left(w-\tfrac{1}{2}\right)^2 + u^2}, \\
\theta_1 &= 2\tan^{-1}{\frac{y}{|\frac{1}{2}-x|}} =2 \cos^{-1}{\frac{y}{r}}, \\
\theta_2 &= 2\tan^{-1}{\frac{u}{|\frac{1}{2}-w|}} =2 \cos^{-1}{\frac{u}{r}}.
\end{split}
\end{equation}
With the exception of the singular, maximally mixed, $r=0$ cases, these functions complete the one-to-one map between column-vector entries, geometric parameters, and toroid statepoints. For the maximally mixed cases, as shown above, there is a one-to-one map between column-vector entries and torus knots. The annular toroid thus has a unique point or knot for each unique pure 2-qubit state in the real subspace.


\begin{acknowledgments}
The authors wish to thank Hal Haggard and Ian Durham for their helpful comments on the manuscript. We gratefully acknowledge the Bard Summer Research Institute and the Bard Office of Undergraduate Research for supporting this work. 
\end{acknowledgments}

The authors have no conflicts to disclose.

\end{document}